\documentclass[twocolumn,showpacs,amsmath,amssymb]{revtex4}
\usepackage{graphicx}
\usepackage{dcolumn}
\usepackage{bm}
\usepackage{amsmath}
\usepackage{amsfonts}
\usepackage{epsfig}
\begin{document}

\title{Enhancement of dark matter capture by neutron stars in binary systems}
\author{Lionel {\sc Brayeur}}\email{lbrayeur@vub.ac.be}
\author{Peter {\sc Tinyakov}}\email{Petr.Tiniakov@ulb.ac.be}
 \affiliation{Service de Physique Th\'eorique,  Universit\'e Libre de Bruxelles, 1050 Brussels, Belgium}

\begin{abstract}
We study the capture of dark matter particles by neutron stars in close binary
systems. By performing a direct numerical simulation, we find that there is a
sizeable amplification of the rate of dark matter capture by each of the
companions. In case of the binary pulsar PSR~J1906+0746 with the orbital
period of 4 hours the amplification factor is $\simeq 3.5$. This amplification
can be attributed to the energy loss by dark matter particles resulting from
their gravitational scattering off moving companions.
\end{abstract}


\maketitle

\section{Introduction}
\label{sec:introduction}

If dark matter (DM) particles interact with nucleons they must accumulate in
stars \cite{Press:1985ug,Gould:1987ju}. Their subsequent annihilation or
collapse into a black hole inside the star may produce observable effects
whose non-detection may be used to impose constraints on dark matter
properties.

In ordinary stars these effects are difficult to detect. Given the direct
constraints on the DM-to-nucleon cross section
\cite{Ahmed:2010wy,Aprile:2011hi}, the DM capture rate in ordinary stars is
too low to eventually form a black hole. Likewise, the heat produced in
annihilations of DM particles is by many orders of magnitude smaller than that
resulting from the nuclear reactions. Only if the annihilation of DM particles
goes into neutrinos, the ones produced in the Sun can be in principle observed
by the neutrino detectors \cite{Jungman:1994jr,Nussinov:2009ft,Menon:2009qj}.

The accumulation of DM is more efficient in compact objects like neutron stars
and white dwarfs. Unlike ordinary stars, these have no internal heat sources
and the heat produced by DM annihilations can, in principle, be detected
\cite{Goldman:1989nd,Kouvaris:2007ay,Sandin:2008db,Bertone:2007ae,McCullough:2010ai,Kouvaris:2010vv,Kouvaris:2010jy}.
Another possible observable consequence of the dark matter accumulation in
compact stars is its eventual collapse into a black hole
\cite{Goldman:1989nd,deLavallaz:2010wp,Kouvaris:2010jy,Kouvaris:2011fi,McDermott:2011jp}.  Both
the DM annihilation and collapse depend crucially on the amount/rate of the DM
captured by the star, so an increase in the capture rate is a potentially
important factor which may result in stronger constraints on the DM
properties.

In order to be captured, a DM particle has to lose a part of its energy and
become gravitationally bound to the star. This energy loss may result from the
DM scattering off the star nucleons. For this mechanism to work, a DM particle
has to cross the star surface. Another mechanism is operative in binary
systems where the gravitational field is time-dependent. In this case DM
particles may lose their energy by means of the gravitational interaction
without crossing the star surface. In the context of the Solar system this
phenomenon, known as the gravitational slingshot, is routinely used to
accelerate/decelerate spacecraft of interplanetary missions. This mechanism
affects larger number of particles than the scattering off the
nucleons. Deceleration of DM particles by moving companions of binary
systems may, therefore, increase the number of gravitationally bound DM
particles and amplify their capture rate.

In the Solar system the effect of the planets (notably, Jupiter) on the dark
matter capture by the Sun has been quantitatively studied in
Refs.\cite{Lundberg:2004dn,Peter:2009mi,Peter:2009mm} and was shown to be small. The
smallness is due to the combination of two factors: (i) the inverse process
(acceleration of DM particles by the gravitational slingshot) prevents a
substantial accumulation of DM and (ii) the Jupiter velocity is much smaller
than that of DM particles, so only a tiny part of the DM phase space is
affected.

Both arguments are not directly applicable to close binary systems involving
neutron stars. While the re-acceleration of DM particles still prevents an
infinite DM accumulation, this effect may be less important in view of a
higher capture rate by neutron stars as compared to the Sun. On the other
hand, the velocities of companions are comparable to those of DM particles, so
that the whole phase space of DM is affected. Thus, an unsuppressed effect of
the companion motion on the DM capture rate  in such systems may be expected.

In this paper we calculated numerically the amplification of DM capture rate in
close binary systems involving neutron stars. As a prototype we considered the
binary pulsar PSR~J1906+0746 composed of two neutron stars of approximately
1.3$~M_\odot$ with the period of orbital motion of 4 hours. In this system the
velocity of the companions is $270$~km/s, which is comparable to a typical
velocity of DM particles in the Galaxy.

We have found that in this binary system each of the companions captures DM at
a rate $\sim 3.5$ times higher than if it were an isolated star. The
enhancement can be attributed to the combination of two factors: the velocity
of the star relative to the DM distribution, and the scattering off the moving
companions, i.e., the gravitational slingshot. One can see that the first
factor, calculable analytically, is smaller than one and for the above binary
system equals $0.57$. Thus, the amplification due to the gravitational
slingshot alone is about $\sim 6$. We also found that the amplification factor
decreases when the orbital velocity of the companions becomes much smaller
than the typical velocity of the DM particles, in agreement with earlier
studies.

The paper is organized as follows. In Sect.~\ref{sec:numerical-procedure} we
describe the numerical procedure used to calculate the dark matter capture
rate in a binary system. In Sect.~\ref{sec:results-conclusions} we present our
results and conclusions.

\section{Numerical procedure}
\label{sec:numerical-procedure}

Since the velocities of DM particles and neutron stars are
comparable, for most of the particle trajectories the interaction with both
companions has to be taken into account. No analytic approximation seems
possible in this case, and particle trajectories has to be calculated
numerically.

In principle, the numerical calculation of the capture rate is
straightforward: one injects particles according to the Maxwellian
distribution, traces them until they either cross the surface of the neutron
star and get captured or leave to infinity, and determines the capture rate
from the fraction of captured particles. In practice, a substantial gain in
calculational time is achieved by treating some parts of this process
analytically.

Consider first the motion of the binary system itself. For simplicity, we
assume that it is composed of two neutron stars of equal mass $M$ moving on a
circular orbit. From the Newton equations one finds
\begin{equation}
{G_NM\over 4R^3\omega^2}=1,
\label{eq:units}
\end{equation}
where $R$ is the radius of the orbit and $\omega$ is the frequency of the
orbital motion of the neutron stars.  The motion of a DM particle in the
vicinity of this binary system is most naturally described in units where
lengths are measured in units of $R$ and times are measured in unit of
$1/\omega$. In these units the velocity of the neutron stars is $v = \omega R
=1$. Assuming the parameters of the binary pulsar PSR~J1906+0746, in the above
units the velocity of light is $c=1127 \omega R$, while a typical velocity of
DM particles in the Galactic halo is $v_{\rm DM}\sim 270\, {\rm km/s} \simeq
\omega R$. Note that the radius of the neutron stars is much smaller than $R$,
\begin{equation}
R_{\rm NS} \simeq 12\, {\rm km} \simeq 2\times 10^{-5}R.
\label{eq:NSradius}
\end{equation}

Let us now turn to particle trajectories. Defining two distance scales
$R_{\rm max} \gg R$ and $R_0\ll R$, we separate each trajectory in
three different regions: (i) the exterior of the sphere $r=R_{\rm
  max}$ which we will refer to as the injection sphere (ii) the small
spherical regions of the radius $R_0$ around the neutron stars, and
(iii) the interior of the injection sphere excluding the vicinities of
the neutron stars.

In the region (i) the binary system may be approximated by a single
central mass $2M$, so the particle equations of motion can be solved
analytically.  Given the asymptotic velocity $v_\infty$ and the impact
parameter $\rho$, the parameters of the particle trajectory at
$r=R_{\rm max}$ are
\begin{equation}
v = \sqrt{v_\infty^2 + {4GM\over R_{max}}},
\label{eq:v}
\end{equation}
\begin{equation}
\sin\psi = {\rho v_\infty\over R_{max} v},
\label{eq:psi}
\end{equation}
where $v$ is the particle velocity and $\psi$ is the angle between the
velocity and the internal normal to the injection sphere.  We generate
the asymptotic parameters $v_\infty$ and $\rho$ according to the
distribution
\begin{equation}
dF \propto \rho  v_\infty^3 \exp \left( - { 3 v_\infty^2 \over 
2 v_{DM}^2 }\right) \, d\rho dv_\infty.
\label{eq:asym_distro}
\end{equation}
The extra power of $v_\infty$ appears here because we are interested in flux
rather than the density of particles.

For each generated $v_\infty$ and $\rho$ we then calculate the parameters of
the trajectory $v$ and $\psi$ at the boundary of the injection sphere
$r=R_{\rm max}$. These parameters should be supplemented with the position of
the entry point (two parameters) and the orientation of the projection of the
velocity in the plane tangent to the sphere. We choose these additional
parameters randomly. The resulting set of parameters fixes the initial
conditions for propagation of the particle trajectory inside region (iii).

Inside the injection sphere the spherical symmetry is no longer a good
approximation. In this region we propagate the particle trajectories
numerically until they either escape outside (beyond $200R$) with positive
energy or get within the distance $R_0$ from one of the neutron stars.

In the region (ii), i.e., close to one of the neutron stars, the
gravitational force of the remote companion can be neglected, and the problem
once again becomes tractable analytically. Taking the initial conditions from
numerical simulations, the trajectory can be continued inside $r=R_0$ and the
point of the closest approach to the neutron star can be determined. It is 
expressed in terms of the particle energy per mass $E$ and the angular momentum
per mass $J$ in the reference frame where the star is at rest, 
\[
R_{\rm min} = {J^2\over 4} \left\{1+\sqrt{1+EJ^2/8}\right\}^{-1},
\]
where the units $R=\omega=1$ were used. In this way the fraction of particles
that get closer than a certain distance to one of the neutron stars can be
computed.

Since the probability for a particle to actually cross the neutron star
surface in numerical simulations is very small in view of the small value of
$R_{\rm NS}$, eq.~(\ref{eq:NSradius}), it is not realistic to measure the
fraction of such particles directly with a reasonable accuracy. The problem
can be avoided by noting that in the gravitational potential of a single star
the cross section for crossing a sphere of radius $R_s$ is equal to
\begin{equation}
\sigma(R_s) = \pi R_s^2 \left( 1 + 
{2 G_N M \over R_s v_{\rm as}^2}\right),
\label{eq:crossection}
\end{equation} 
where $v_{\rm as}$ is the asymptotic velocity of a particle. For the actual
parameters of the neutron star system the second term dominates when $R_s\ll
R$. One thus expects that for a sufficiently small $R_s$ the cross section of
crossing the sphere $r=R_s$ scales linearly with $R_s$. Since the dependence
on the velocity factorizes in this regime, the linear behavior with $R_s$ is
maintained after averaging over particle velocities. Measuring the fraction
of particles crossing $r=R_s$ at different values of $R_s$ small enough for
the linear scaling, one may then extrapolate to the actual neutron star
radius.

We note in passing that eq.~(\ref{eq:crossection}) also explains why the
capture rate by a single neutron star decreases with the increasing relative
velocity of the star and dark matter: with the first term inside the brackets
on the r.h.s. neglected, the cross section is proportional to $1/v_{\rm
  as}^2$. Depending on the direction of the particle velocity the effect of
the neutron star motion can have both signs. Upon averaging over the
Maxwellian distribution the net result is the decrease of the capture rate
with the increasing velocity of the neutron star.

\section{Results and conclusions}
\label{sec:results-conclusions}

In our simulations we did not fix the absolute normalization of the
distribution of injected particles. Instead, for each value of the parameters
we performed two identical simulations: one with moving neutron stars and one
with the neutron stars fixed to their positions at a given moment of time. In
the second case all the effects due to the star motion disappear, and the
capture rate equals, to a good approximation, twice the capture rate of a
single isolated neutron star. The amplification factor thus equals the ratio
of the number of particles captured in these two simulations. Clearly, this
ratio does not depend on the absolute normalization of particle distributions.

\begin{figure}[h]
\begin{center}
\includegraphics[width=\columnwidth]{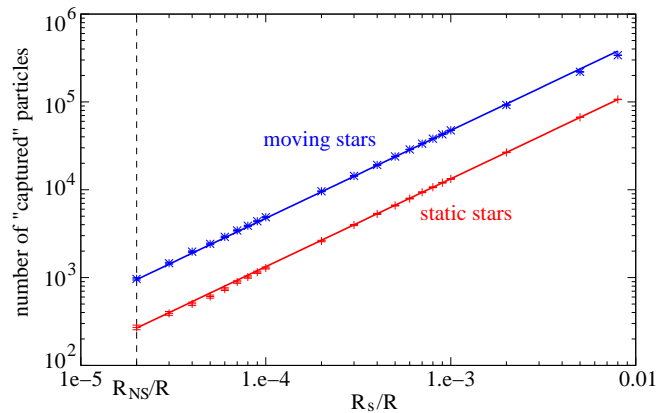}
\end{center}
\caption{The dependence of the number of ``captured'' particles (i.e., those
  crossing the sphere $r=R_s$) on the capture radius $R_s$ for moving (upper
  line) and static (lower line) neutron stars. The points represent numerical
  simulation with 1-$\sigma$ statistical errors. The lines show fits with the
  linear dependence. The actual radius of the neutron star is shown by the
  vertical dashed line. The number of injected particles was
  $10^8$.\label{fig:Rs} }
\end{figure}

In each of the two simulations we measured the number of particles that got
closer than $R_s$ to one of the neutron stars, as a function of $R_s$. The
results are presented in Fig.~\ref{fig:Rs}. The parameters of the simulation
were as follows: $R_{\rm max} =10R$, $R_0=10^{-3} R$, and the number of thrown
particles was $10^8$. One can see that the expected linear dependence with
$R_s$ is well reproduced. Fitting both dependencies with a linear function and
taking the ratio of the two coefficients we find that the amplification factor
equals $3.5\pm 0.1$. 

The amplification factor depends on the parameters of the binary system, as it
certainly has to go to one for slowly moving stars. The neutron star motion
(assuming stars of equal mass) is determined by their mass and rotation
period. When the trajectories of DM particles are expressed in the units of
$R$ and $1/\omega$, the change of the binary system parameters translates into
the change of asymptotic velocity of the DM particles expressed in terms of
$R\omega$, and the value of the capture radius that corresponds to the actual
radius of the neutron star. As we have seen, the latter parameter is
irrelevant. Thus, the only important parameter is the ratio of the asymptotic
velocity of DM to that of the neutron stars. Equal values of this parameter
imply equal amplification factors.

We have calculated the amplification factors for several values of the binary
system periods, assuming the masses of the companions equal $M=1.3M_\odot$.
The results are presented in Table.~\ref{tab:amplification_of_T}. The maximum
amplification corresponds to the case when the velocities of the companions and
the asymptotic velocity of dark matter particles are comparable, while when
the neutron stars move much slower than the DM particles, the amplification
factor approaches 1. The amplification factor also decreases for short periods
when the neutron star velocity becomes larger than that of DM particles,
because the suppression due to the motion of the neutron stars relative to
dark matter (cf. eq.~(\ref{eq:crossection})) is no more compensated by the
gain in the phase space of DM affected by the
gravitational slingshot effect.

\begin{table}[h]
\begin{tabular}{c|c}
\mbox{period} & amplification \\\hline
{\mbox 4h} & 3.5\\
{\mbox 8h} & 4.3\\
{\mbox 16h} & 2.8\\
{\mbox 32h} & 1.5
\end{tabular}
\caption{\label{tab:amplification_of_T} The dependence of the amplification
  factor on the period of the binary system. }
\end{table}

To summarize, we have shown that the motion of the neutron stars in close
binaries leads to the amplification of the DM capture rate by a factor of up
to 3--4 depending on the period of the binary system. Thus, binary systems are
favorable place to put constraints on dark matter properties that follow from
its capture in compact stars.

\section*{Acknowledgments}
This work is supported in part by the IISN project No. 4.4509.10, by the ARC
project “Beyond Einstein: fundamental aspects of gravitational interactions”
and by the Russian Federation Ministry of Education under Contract
14.740.11.0890.

\end{document}